\documentclass[aps,prb,showpacs,twocolumn,reprint,superscriptaddress]{revtex4-1}
\usepackage{amsmath}
\usepackage{amssymb}
\usepackage{graphicx}
\usepackage{color}
\usepackage{multirow}
\usepackage{dcolumn}
\usepackage{textcomp}
\usepackage{longtable}
\usepackage{makecell}
\begin{document}

\title{Electrically switchable valley polarization, spin/valley filter, and valve effects in transition-metal dichalcogenide monolayers interfaced with two-dimensional ferromagnetic semiconductors
}

\author {Ao Zhang}

\author{Kaike Yang}
\affiliation{School of Physics and Electronics, Hunan Normal University,
Key Laboratory for Matter Microstructure and Function of Hunan Province,
Key Laboratory of Low-Dimensional Quantum Structures and Quantum Control of Ministry of Education, Changsha 410081, China}

\author {Yun Zhang}
\affiliation{Baoji University of Arts and Sciences, Baoji 721016, China
}

\author {Anlian Pan}
\affiliation{Key Laboratory for Micro-Nano Physics and Technology of Hunan Province, 
College of Materials Science and Engineering, Hunan University, Changsha 410082, China
}
\author {Mingxing Chen}
\email{mxchen@hunnu.edu.cn}
\affiliation{School of Physics and Electronics, Hunan Normal University,
Key Laboratory for Matter Microstructure and Function of Hunan Province,
Key Laboratory of Low-Dimensional Quantum Structures and Quantum Control of Ministry of Education, Changsha 410081, China}

\date{\today}

\begin{abstract}
 Electron valleys in transition-metal dichalcogenide monolayers drive novel physics and allow designing multifunctional architectures for applications. We propose to manipulate the electron valleys in these systems for spin/valley filter and valve devices through band engineering. Instead of the magnetic proximity effect that has been extensively used in previous studies, in our strategy, the electron valleys are directly coupled to the spin-polarized states of the two-dimensional ferromagnets. We find that this coupling results in a valley-selective gap opening due to the spin-momentum locking in the transition-metal dichalcogenide monolayers. This physics gives rise to a variety of unexpected electronic properties and phenomena including halfmetallicity, electrically switchable valley polarization, spin/valley filter and valve effects in the transition-metal dichalcogenide monolayers. We further demonstrate our idea in MoTe$_2$/CoCl$_2$ and CoCl$_2$/MoTe$_2$/CoCl$_2$ van der Waals heterojunctions based on first-principles calculations. Thus, our study provides a way of engineering the electron valleys in transition-metal dichalcogenide monolayers for new-concept devices.
\end{abstract}

\keywords{Interface; Transition-metal dichalcogenide monolayer; valleytronics; band structure}

\maketitle

Electron valleys in solids have led to the emergence of the so-called valleytronics.
Transition-metal dichalcogenide (TMD) monolayers are promising in valleytronics for their unique electronic properties related to the electron valleys
\cite{PRL_valley_2012,Feng_NC_2012,Nat_Rev_Mat_2016}.
These systems have a direct band gap with the valleys located at two inequivalent high symmetry points K and K$^\prime$, which are degenerate due to the time-reversal symmetry.
The broken inversion symmetry and spin-orbit coupling result in a valley-dependent spin-momentum locking, giving rise to many novel physics phenomena, such as valley-selective photon excitation and valley Hall effect\cite{PRL_valley_2012,Feng_NC_2012,Nat_Rev_Mat_2016,VHE_2014}. 

To date, much attention has been paid to valley polarization in TMD monolayers, which is crucial for their applications in valleytronic  devices\cite{Mak_2012,Zeng_2012,Electrical_2017}. 
For such a purpose, applying external magnetic fields is an intuitive way since the magnetic fields break the time-reversal symmetry
\cite{Li_Zeeman_2014,Srivastava_2015,Aivazian_2015,MacNeill_2015}. 
Alternatively, one can interface the TMD monolayers with a magnetic substrate. 
In this way, the magnetic proximity effect induced by the substrate causes a Zeeman-type splitting in the electron valleys
\cite{Qi_2016,Cheng_2016,Song_2016,Zhao_2017,Liang_2017,Song_2017,Xu_2018,Yang_2018,Xue_2019,Norden_2019,Zhou_2019,Zhou_2020}. 
Moreover, this type of heterostructures allows nonvolatile valley polarizations whose values can be much larger than those induced by the external magnetic fields.
Recently, the investigation has been extended to van der Waals (vdW) heterostructures, in which a two-dimensional (2D) ferromagnetic (FM) semiconductor serves as the substrate\cite{WSe2_CrI3_Sci_Adv,Xie_2018,Seyler_2018,Zhang_2019,Tang_2020,Ciorciaro_2020,Marfoua_2020,Hu_PRB_2020,Mak_2020}. 
The valley polarization can be tuned by applying gate voltages and even be switched if the magnetization of the substrate is reversed
\cite{Hu_PRB_2020,Mak_2020}.
These advances have demonstrated the tunability of the electron valleys in the TMD monolayers toward valleytronics.  

Here, we propose a different strategy of engineering the electron valleys in the TMD monolayers for spin/valley filter and valve.
In contrast to previous studies that make use of the magnetic proximity effect, 
in our strategy, the electron valleys are directly coupled to a spin-polarized band of the 2D ferromagnet in TMD/FM vdW heterojunctions. 
This coupling results in a valley-selective gap opening due to the strong spin-valley coupling.
Such physics can lead to halfmetallicity in the TMD monolayer for spin/valley filter and an in-plane spin/valley valve effects if it is sandwiched in between two ferromagnetic monolayers, i.e., FM/TMD/FM junctions.
Based on first-principles calculations, we find that these effects may be realized in vdW heterojunctions made of MoTe$_2$ and CoCl$_2$ monolayers.



\begin{figure*}
  \includegraphics[width=.95\linewidth]{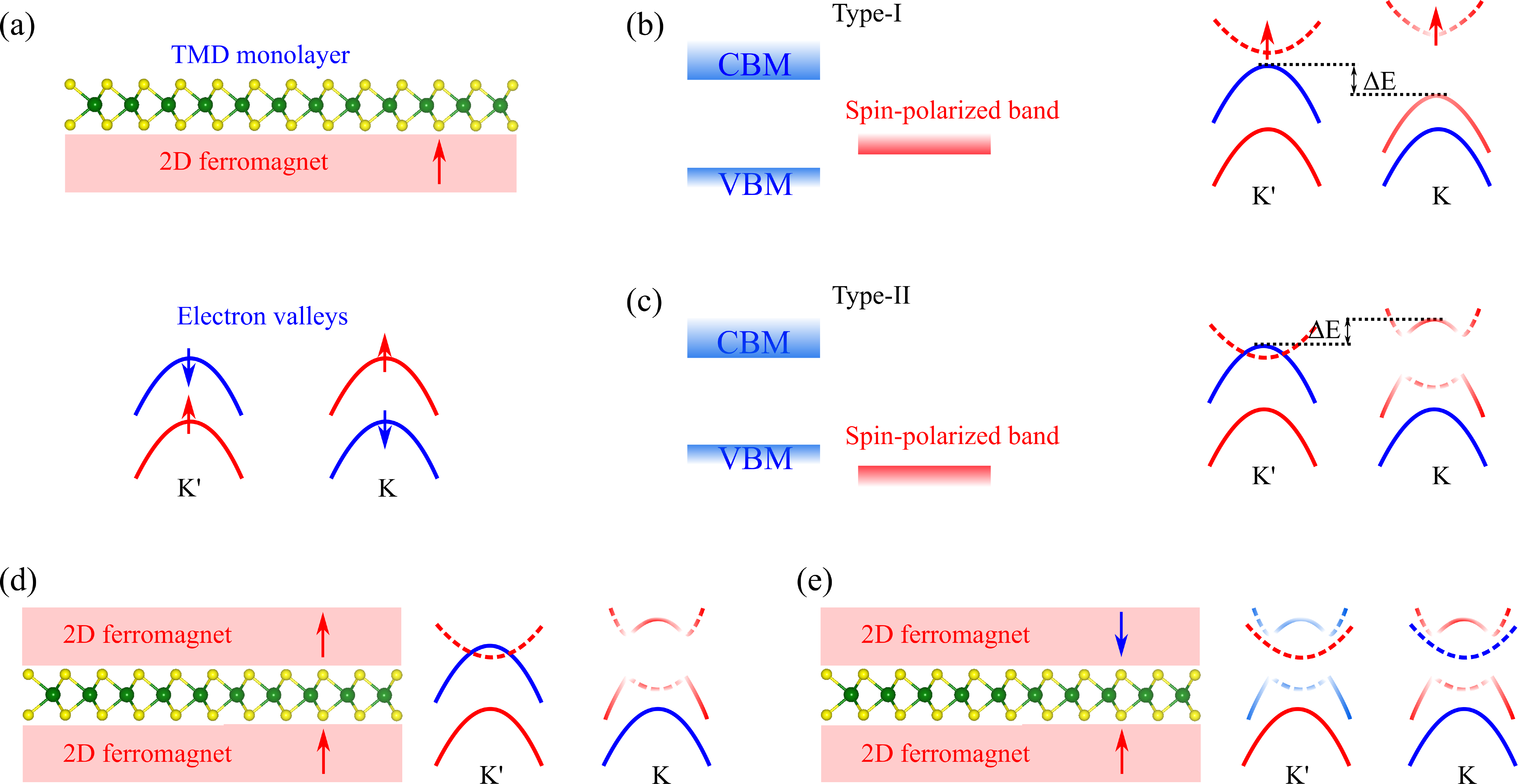}
  \caption{Coupling of a TMD monolayer to a 2D ferromagnet. 
  (a) Schematic illustrations of the heterostructure (upper panel) and the electron valleys in the TMD  monolayers (lower panel). 
  (b,c) Schematic illustrations of the band alignments of the heterostructure (left) and the band structures due to the hybridization between the states of the constituents (right).
  The dashed lines represent the spin-polarized bands of the 2D ferromagnet.
  There are two types of band alignments between the valence band maximum (VBM) of the TMD monolayer and the spin-polarized band of the substrate.
  The spin-polarized band of the FM monolayer can lie either higher (b) or lower (c) than the VBM of the TMD monolayer, which are defined as type-I and type-II alignments, respectively.
  Arrows show the spins of the electronic bands.
  $\Delta E$ denotes the valley polarization in the TMD monolayer, i.e., $\Delta E = E_{VBM}(K') - E_{VBM}(K)$.
  (d,e) In-plane spin/valley valve effect in a FM/TMD/FM junction.}
 \label{fig1}
\end{figure*}
We illustrate our idea in Fig.~\ref{fig1}. We focus on the valence bands of the TMD monolayer since there is a large spin splitting at K (K$^\prime$) due to the spin-orbit (SO) coupling\cite{TMD_PRB_2011}.
The two valleys have opposite spin states due to the time-reversal symmetry.
We assume that a FM monolayer has a spin-polarized band that lies nearby the VBM of the TMD monolayer when their bands align, which can be higher or lower than the K and K$^\prime$ valleys of the TMD monolayer.
We refer to them as type-I and type-II band alignments, respectively.
We further assume that this band is parabolic around K (K$^\prime$) with the same spin as the K valley.
Because the spin-up and spin-down components of the valence band of the TMD monolayer at K (K$^\prime$) are completely decoupled\cite{PRL_valley_2012},
it is safe to assume that only the spin-up component of the valence band is coupled to the spin-polarized band of the 2D ferromagnet near the K point when their bands align properly.
The band hybridization between them is denoted by $\gamma$.
The low-energy effective Hamiltonian for the heterostructure near K can be written as 
\begin{equation}
H(\vec{k})=  
\left[
 \begin{array}{ccc}
     \frac{(\hbar k)^2}{2m_1^*} & 0                                & \gamma \\
     0                          & \frac{(\hbar k)^2}{2m_1^*}-\Delta_{SO} & 0      \\
     \gamma*                    & 0      & \frac{(\hbar k)^2}{2m_2^*} + \mu 
 \end{array}
 \right] ,
 \label{eqH}
\end{equation}
where $\Delta_{SO}$ denotes the SO splitting in the valence band of the TMD monolayer,
and $\mu$ represents the chemical potential of the 2D ferromagnet, which can be tuned by external electric fields. 
Correspondingly, the band alignment may be changed.
We have also assumed that the effective mass $m_2^*$ has an opposite sign to $m_1^*$, which is approximated by $m_1^* = -\frac{\hbar^2}{2Ja^2}$ with $a$ and $J$ being the lattice constant and hopping parameter between the nearest neighbors. The eigenvalues are given in the unit of $J$ and $\mu/J$ and $\gamma/J$ serve as parameters.
The K valley will be pushed down to a lower (higher) energy for type-I (type-II) band alignment as a result of $\gamma$.
Whereas, the K$^\prime$ valley remains unchanged since it has a different spin state from the spin-polarized band (Fig.~\ref{fig1}).
This valley-selective coupling leads to a valley polarization, i.e., the degeneracy of the electron valleys is removed. If we take the K valley as the reference, then the valley polarization is positive for type-I band alignment and becomes negative  for type-II band alignment.
Therefore, one can obtain a switchable valley polarization across the band alignment transition, which can be realized by applying external electric fields instead of magnetic fields.

Note that in the case of the type-II band alignment, the hybridization opens up a gap at the K valley (Fig.~\ref{fig1}c). 
Therefore, one can expect that sandwiching a TMD monolayer in between two FM monolayers can give rise to a spin/valley valve effect: one valley is metallic/halfmetallic and the other one is semiconducting when the two FM layers have the same magnetization. 
Reversing the magnetization of the top FM layer results in gap openings at both valleys (see Fig.~\ref{fig1}e). 


The above picture requires that a suitable FM monolayer should satisfy the following criteria:
(i) it has a vdW-type interaction with the TMD monolayer so that the electron valleys can be preserved well. Otherwise, the electron valleys will be strongly disturbed by the strong bonding with the substrates\cite{AoZhang_2020}.
(ii) It has a proper work function such that its bands at K and K$^\prime$ lie nearby the VBM of the TMD monolayer when their bands align.  
The newly discovered 2D FM semiconductors such as CrI$_3$ and CrGeTe$_3$ may be candidates
\cite{CrI3_Nature,Nature_CrGeTe3,CrI3_Zhang}. 
However, previous studies revealed that there is a large energy difference ($>$ 0.5 eV) between the bands of a CrI$_3$ as well as CrGeTe$_3$ monolayer and a WSe$_2$ monolayer at K (K$^\prime$), which gives rise to a small valley polarization ($\sim$ 3 meV).

We find that a CoCl$_2$ monolayer may be a good candidate for the FM substrate in the interface structure with a MoTe$_2$ monolayer\cite{CoCl2_2019}. 
First, it has a layered structure that may be obtained from its bulk phase\cite{CoCl2_2019} 
Next, it has a small lattice mismatch with MoTe$_2$ ($<$ 2\%). These features are good for this system to form a vdW heterostructure with the MoTe$_2$ monolayer. Moreover, we find that its conduction band, spin-polarized and largely separated from other bands, lies slightly lower than the VBM of the MoTe$_2$ monolayer at K, which is desired for our model.

\begin{figure*}
\centering
\includegraphics[width=.89\linewidth]{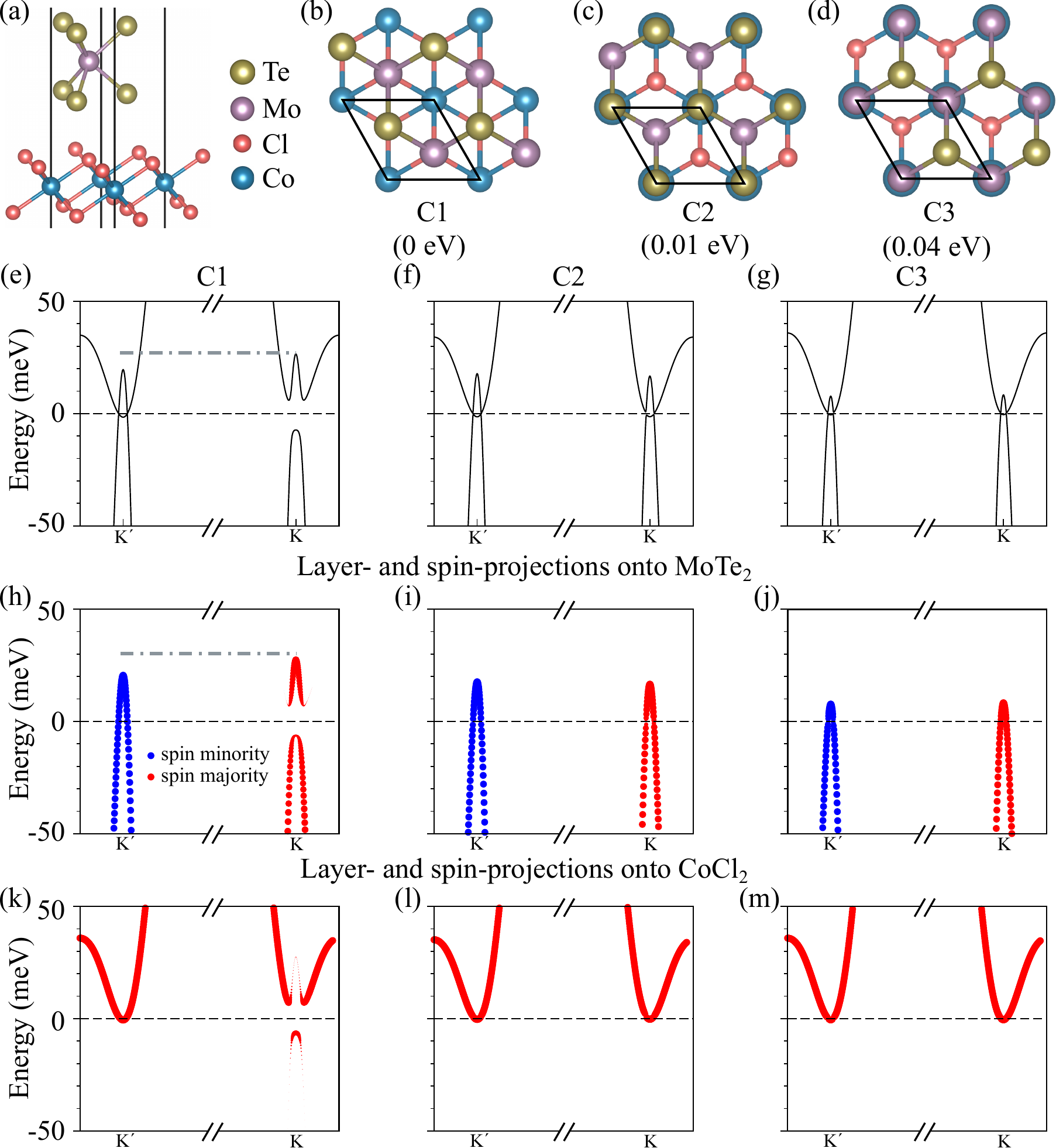}
\caption{Valley-selective gap opening in MoTe$_2$/CoCl$_2$. 
  (a) Perspetive view of the heterostructure. 
  (b)-(d) Top views of the heterostructure in three high-symmetry stackings and (e)-(g) corresponding band structures. 
  (h)-(m) Layer- and spin-projected band structures for the three configurations.
  The energy difference relative to the lowest energy structure (C1) is given below each configuration.
  The dash-dotted lines in (e) and (h) are drawn to guide the eye.
  The Fermi level is set to be zero. }
\label{fig2}
\end{figure*}

We have performed density-functional theory (DFT) calculations for MoTe$_2$/CoCl$_2$ and CoCl$_2$/MoTe$_2$/CoCl$_2$ vdW junctions using the Vienna Ab initio Simulation Package\cite{kresse1996}.
We use a slab structure to model the heterostructures and junctions, in which both constituents are in the 1 $\times$ 1 primitive cell.
The pseudopotentials were constructed by the projector augmented wave method\cite{bloechl1994,kresse1999}.
Van der Waals dispersion forces between the adsorbate and the substrate 
were accounted for through the optPBE-vdW functional by using the vdW-DF method 
developed by Klime\v{s} and Michaelides\cite{klimes2010,klimes2011}.
A 15 $\times$ 15 Monkhorst-Pack $k$-mesh was used to sample the two-dimensional Brillouin zone and a plane-wave energy cut off of 320 eV was used for structural relaxation and electronic structure calculations. 
Layer-projection calculations were performed using the KPROJ program\cite{KPROJ}.

We have considered many configurations of MoTe$_2$/CoCl$_2$ since the interaction between them can be dependent on specific stackings. We note that there are three high-symmetry stackings (see Fig.~\ref{fig2}). Other configurations obtained by artificially translating the TMD monolayer with respect to configuration C1 are shown in Fig.~S1 (see the Supplemental Material (SM)\cite{SM}). Our calculations find that C1 has the lowest energy. The overlayer and substrate interact via a vdW-type binding, for which the binding energy is about 0.2 eV, smaller than that for WSe$_2$/CrI$_3$\cite{Xie_2018,Hu_PRB_2020} and WSe$_2$/CrGeTe$_3$\cite{Marfoua_2020}. 
In our calculations, we have taken into account the effect of electron correlations related to the partially filled Co-3$d$ orbitals.
We estimated the effective Coulomb interaction ($U$) by using the linear response method
\cite{Cococcioni_2005}, which gives $U$ = 3.15 eV for the Co-3$d$ orbitals in CoCl$_2$ and is in consistent with previous study \cite{CoCl2U}.

One prominent feature of the band structures is the valley-selective gap opening,
that is, a band gap appears near the K valley for MoTe$_2$/CoCl$_2$.
The gap size is about 13 meV for configuration C1, which is nearly vanishing for C2 as well as for C3.
The gap opening also appears for other configurations (see Fig.~S2 in the Supplementary Material\cite{SM}).
The valley-selective gap opening results in halfmetallicity in MoTe$_2$ for C1 as revealed by the spin-projected band structure (Figs.~\ref{fig2} and S3\cite{SM}).
This feature is unexpected since both the overlayer and substrate are semiconductors.
However, it can be understood with the help of the band alignment since the spin-polarized conduction band of CoCl$_2$ and the VBM of MoTe$_2$ have a band alignment similar to the type-II band alignment shown in Fig.~\ref{fig1}c (also see Fig.~S4 in the Supplementary Material\cite{SM}).
The hybridization between them leads to the gap opening at K.
The valley-selective gap opening also leads to valley polarization.
The reason is that the VBM of MoTe$_2$ at K is pushed up to a higher energy than that at K$^\prime$ during the formation of the hybridization gap.
Therefore, the magnitude of the valley polarization depends on the gap opening: C1 shows a valley polarization of about 7 meV, which is much larger than other configurations.

The band alignments of MoTe$_2$ and CoCl$_2$ favor charge transfer. However, our calculations find that the charge transfer between them is only 0.02 \textit{e}, which is comparable to that for WSe$_2$/NiCl$_2$ with a similar band alignment\cite{Tang_2020}. The reason is that they interact via a vdW-type bonding, which instead causes a charge redistribution in CoCl$_2$ (see Fig.~S5 in the Supplementary Material\cite{SM}).

\begin{figure}
\centering
\includegraphics[width=.95\linewidth]{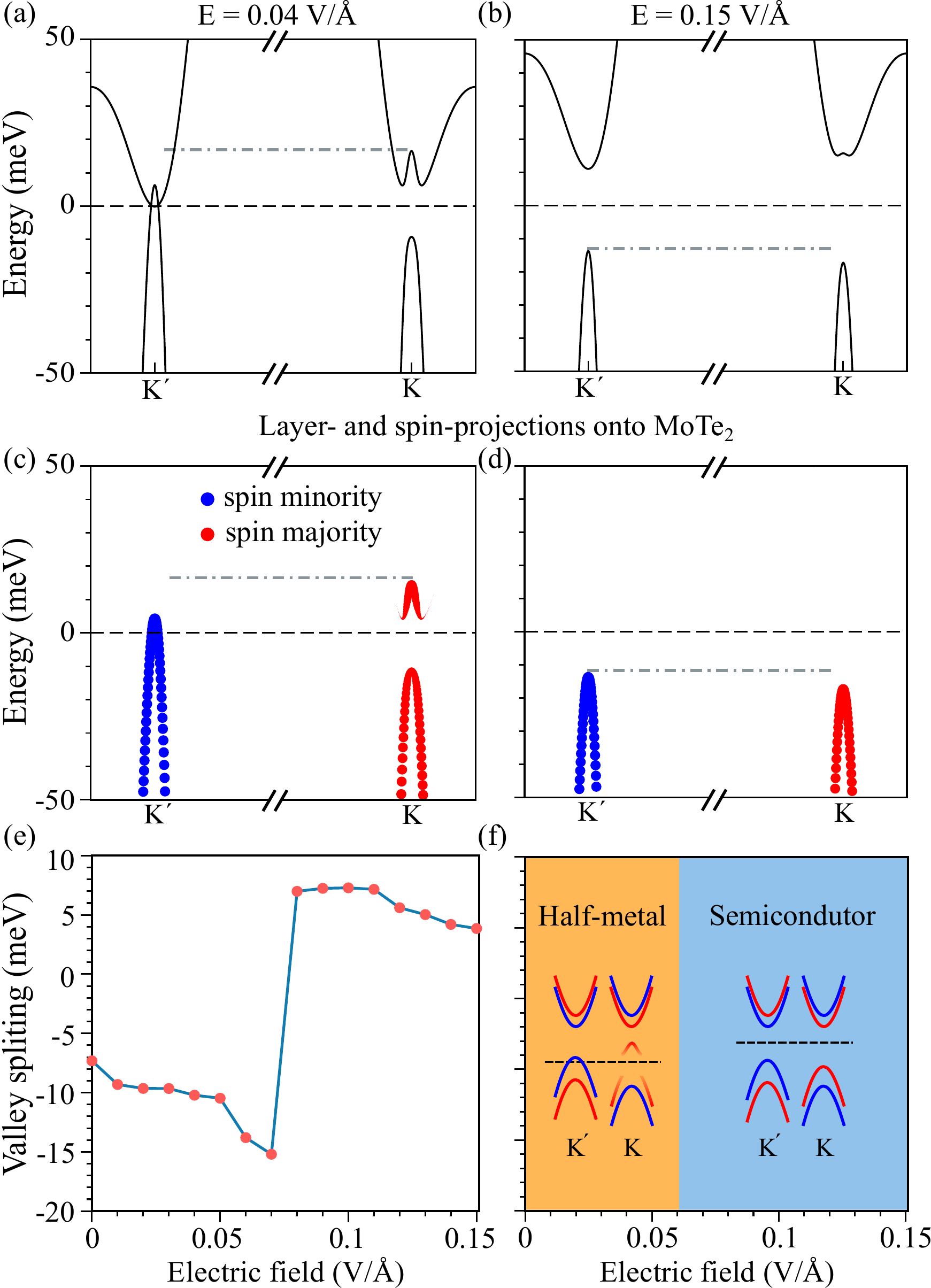}
\caption{Electrical switching of the valley polarization in MoTe$_2$/CoCl$_2$. 
  (a, b) Band structures of the heterostructure under different electric fields.
  (c, d) Corresponding band structures with layer- and spin-projections onto MoTe$_2$.
  The dash-dotted lines in (c) and (d) are used to guide the eye.
  (e) Valley splitting as a function of gate voltage.
  (f) Phase diagram of MoTe$_2$/CoCl$_2$ under external electric fields.}
\label{fig3}
\end{figure}

Our calculations find that interfacing induces magnetic moments on Mo atoms, which are about 0.003 $\mu_B$ for all configurations. However, they show distinct trends in the gap opening and valley polarization at K. Therefore, these behaviors may not be due to the magnetic proximity effect, but band hybridization.
We find that the differences can be understood within the tight binding (TB) approximation.
The conduction band of CoCl$_2$ at K/K$^\prime$ is contributed by the Co-3$d_{3z^2-r^2}$ orbital (see Fig.~S6 in the Supplementary Material\cite{SM}).
Whereas the valence band of MoTe$_2$ at the valleys is contributed by both Mo-4$d_{xy}$ and -4$d_{x^2-y^2}$ orbitals.
The basis function thus is $\phi_{v}^K = \frac{1}{\sqrt{2}}(|d_{x^2-y^2}>+i\tau |d_{xy}>)$, where $\tau = 1$ and $-1$ for the K and K$^\prime$ valleys, respectively\cite{PRL_valley_2012}.  
The hybridization $\gamma$ between MoTe$_2$ and CoCl$_2$ near K (K$^\prime$) in Eq.~\ref{eqH} can be approximated as 
\begin{equation}
\gamma = H_{Mo,Co}^{\vec{k}}= \sum_{NN}e^{i\vec{k}\cdot \vec{R}_{NN}}
<\phi_{v}^K|H|d_{3z^2-r^2}>
 \label{gamma}
\end{equation}
where $\vec{R}_{NN}$ denote the lattice vectors from a Mo atom to its nearest-neighboring Co atoms.
Calculations of Eq.~\ref{gamma} based on the Slater-Kostor method\cite{TB} find that $\gamma$ is finite for C1 and equals to zero for C2 and C3 (see the Supplemental Material and Fig.~S7\cite{SM}).
We derive the Hamiltonian matrix elements in the basis of the Wannier orbitals\cite{wannier90}, which confirm this trend.

Based on the above understanding, one can learn that the halfmetallicity and valley polarization are dependent on the band alignments of the two constituents, which can be tuned by gate voltages.
In Fig.~\ref{fig3}, we show the band structures of C1 under different electric fields (Spin-projected band structures are shown in Fig.~S8\cite{SM}).
The bands of the MoTe$_2$ monolayer move down to lower energies upon positive gate voltages, 
which drives a halfmetal-semiconductor transition in the system. 
The critical value of the electric field for the transition is about 0.07 V/{\AA}.
Note that the valley polarization is switched across the transition.
The valley polarization in our system can be tuned to be one order of magnitude larger than those for WSe$_2$/CrI$_3$ and WSe$_2$/CrGeTe$_3$ as revealed by previous studies
\cite{WSe2_CrI3_Sci_Adv,Xie_2018,Seyler_2018,Zhang_2019,Tang_2020,Ciorciaro_2020,Marfoua_2020,Hu_PRB_2020,Mak_2020}.
We further confirm that the above effects are robust against different $U$ parameters (see Figs.~S9 in the Supplementary Material\cite{SM}).


\begin{figure}
\centering
\includegraphics[width=.90\linewidth]{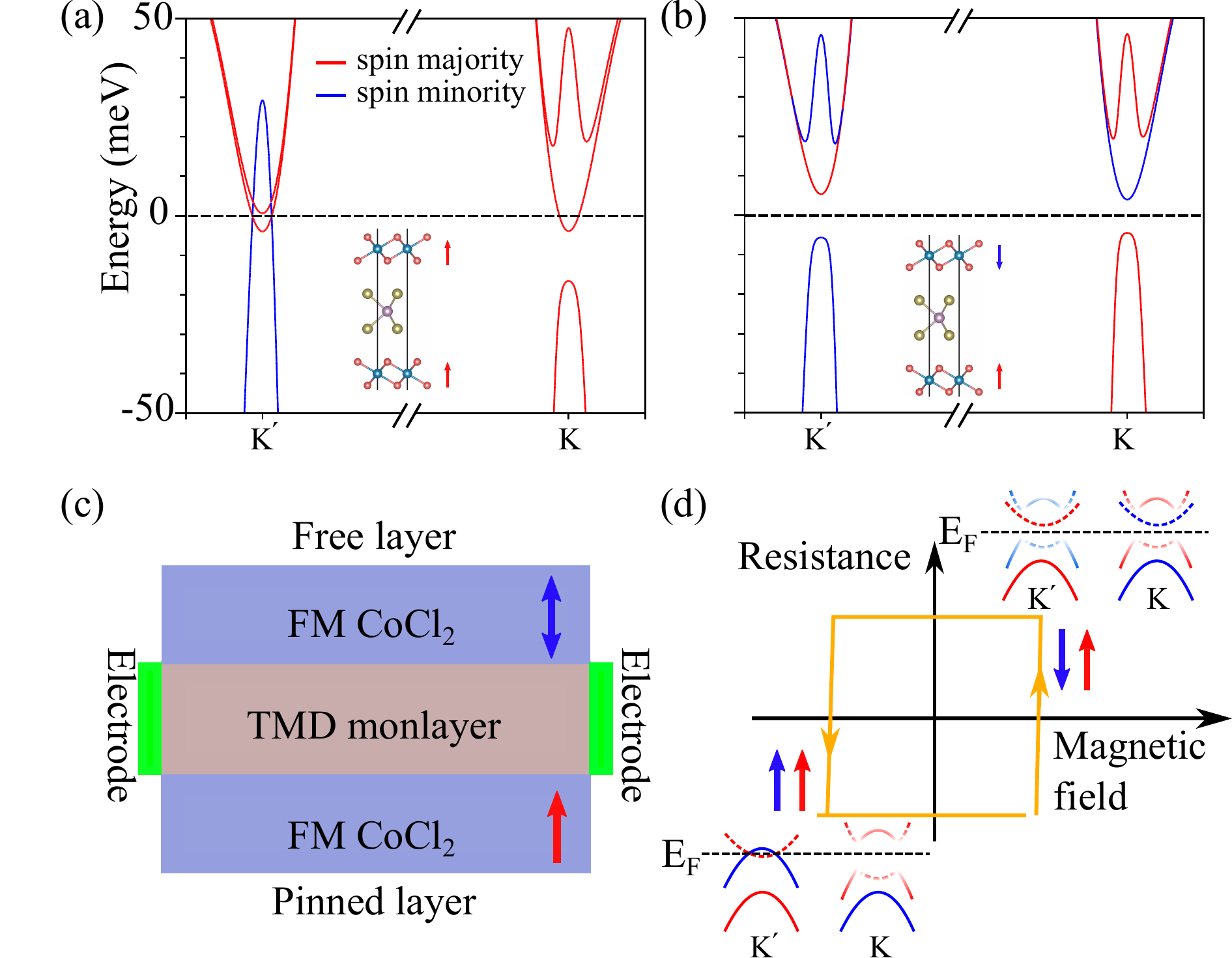}
\caption{Spin/valley valve effect in CoCl$_2$/MoTe$_2$/CoCl$_2$ junction. 
  Spin-projected band structures around K and K$^\prime$ for that the magnetizations of the two CoCl$_2$ layers are aligned ferromagnetically (a) and anti-ferromagnetically (b), respectively. 
  The red and blue lines denote different spins states.
  The insets show the structure of the junction and the magnetizations of the CoCl$_2$ monolayers. The Fermi level is set to be zero. 
  (c, d) Proposed spin and valley valve made of a FM/TMD/FM junction.}
\label{fig4}
\end{figure}
We now proceed to CoCl$_2$/MoTe$_2$/CoCl$_2$ junctions, which are constructed based on the lowest energy configuration C1 shown in Fig.~\ref{fig2}.
The two FM monolayers can behave like the pinned and free layers in the magnetic tunnel junctions (MTJs), respectively.
We assume that the bottom CoCl$_2$ layer serves as the pinned layer, which can be realized by coupling it to either a strong ferromagnetic or antiferromagnetic layer.
The magnetizations of the two CoCl$_2$ monolayers can align either ferromagnetically or anti-ferromagnetically by manipulating the magnetization of the top CoCl$_2$ layer using external magnetic fields.
We have performed calculations for the two kinds of magnetic configurations, for which the results are summarized in Fig.~\ref{fig4}.
As expected, there is only one valley with gap opening when the magnetizations of the CoCl$_2$ are in the FM configuration.
Layer projection indicates that the MoTe$_2$ monolayer remains halfmetallic due to the gap opening at the K valley (see Fig.~S10 in the Supplementary Material\cite{SM}). 
Compared to the heterostructure MoTe$_2$/CoCl$_2$, there is a band contributed by the CoCl$_2$ layers crossing the Fermi level of the junction near the K point.
This band originates from the interaction between states of the two CoCl$_2$ monolayers.
When the magnetizations of the two CoCl$_2$ layers are aligned anti-ferromagnetically by reversing that of the top layer, gap openings occur at both valleys.
Thus, one can achieve in-plane spin as well as valley valve effect in this type of junction.
This type of spin valve is distinctly different from traditional MTJs 
in that our system allows both spin and valley valve simultaneously (Figs.~\ref{fig4}c, d).
Moreover, it is also distinct from what has been experimentally realized in NiFe/MoS$_2$/NiFe\cite{NL_2015}, where the TMD layer plays simply the same role as the MgO layer in the traditional MTJs.

In addition, we would like to discuss the effects of the strong magnetic substrates used to pin the magnetization of the bottom CoCl$_2$ layer on the band structure of CoCl$_2$/MoTe$_2$/CoCl$_2$ junctions. When a metal such as permalloy is used, perturbations over the band structure of the bottom CoCl$_2$ layer can be large since strong band hybridizations between them may exist. However, if one instead uses a magnetic insulator with a proper work function such that the conduction band of the bottom CoCl$_2$ layer lies in its band gap, disturbations over the band of CoCl$_2$ can be negligibly small. This type of architecture is favorable for the spin/valley valve proposed in this study. We anticipate that our results will stimulate experiments to explore the spin/valley filter and valve effects in this type of vdW heterostructures.

In summary, 
we have proposed a strategy of engineering the electron valleys in TMD monolayers for applications in spin/valley filter and valve by interfacing them with 2D ferromagnets.
Our strategy makes use of the spin-momentum locking in the TMD monolayers.
When the valance band of the TMD monolayers aligns properly with a spin-polarized band of a 2D FM monolayer, only one of the valleys is coupled to the spin-polarized band.
This coupling induces a valley-selective gap opening and results in  valley-polarization and halfmetallicity in the TMD monolayer, which can be used for spin/valley filter.
We have further demonstrated that the spin/valley filter effect may be realized in the vdW heterojunction MoTe$_2$/CoCl$_2$ based on first-principles calculations.
Moreover, we find that this type of heterojunctions exhibits a switchable valley polarization and halfmetal-semiconductor transition under external electrical fields.
We have also demonstrated that sandwiching a MoTe$_2$ in between two CoCl$_2$ monolayers, i.e., CoCl$_2$/MoTe$_2$/CoCl$_2$ junction, allows realizing in-plane spin/valley valve effect in the TMD monolayer. 
This type of spin/valley valve that distinctly differs from traditional MTJs 
is expected to have potential applications in new functional devices.  

\begin{acknowledgments}
We acknowledge Wang Yao for fruitful discussions. This work was supported by the National Natural Science Foundation of China (Grants No. 11774084, No. U19A2090, No. 11804333 and No. 91833302).
\end {acknowledgments}

\bibliography{references}
\bibliographystyle{apsrev4-1}

\end{document}